\documentstyle[twocolumn,aps]{revtex}

\begin{document}


\title
{\bf Spontaneous spin polarized tunneling current through
a quantum dot array}

\author{David M.-T. Kuo and Y.-C. Chang}
\address{ Department of Physics and Materials Research Laboratory\\
University of Illinois at Urbana-Champaign, Urbana, Illinois
61801}
\date{\today}
\maketitle

\begin{abstract}
We show theoretically that a strongly spin-polarized current can
be generated in semiconductors by taking advantage of the
ferromagnetic phase of a quantum dot array (QDA). A Hubbard
model with coupling to leads is used to study the tunneling current of
the QDA system as a function of gate voltage. Due to the weak interdot
coupling and strong Coulomb repulsion, it is found that a
ferromagnetic phase exists in QDA within a window of gate voltage.
Therefore QDA can be used as a spin filter to detect
and control spin states in quantum information devices.
\end{abstract}

Recently, spintronics and quantum information (QI) processing have
attracted a great deal of attention.[1] A good QI system should
provide well defined quantum computational space, precise
quantum-state preparation, coherent quantum manipulation, and state
detection[2]. Solid state devices based on modern advanced
semiconductor techniques opened up the possibility of fabricating
large integrated networks which would be required for the
realization of quantum computation[3,4]. Both charged states[5] and
spin states[6] of electrons have been proposed to carry
the quantum information. Because the
decoherence time of spin is much longer than that of charge[6],
using the spin is more promising for quantum information
processing. Kane has proposed to use the nuclear spin as the
quantum bit, since its decoherence time is much longer than that
of the electron spin[7]. Nevertheless, its manipulation is
difficult due to the weak coupling between nuclear spin and
electron spin (hyperfine interaction). Thus, the use of spin
states of electrons in semiconductors remains a viable option. One
of the challenging problems is the preparation of electrons in
semiconductors with well defined spin state.

Controlling electron spin states, such as coherent manipulation
and filtering, becomes crucial in the implementation of quantum
computer. DiVincenzo [8] has suggested to use the spin filter
effect to manipulate spin states.  Semiconductor quantum dots with
local magnetic field can be used as spin filter and momory (read
out) device [9]. Ferromagnetic semiconductor materials can also be
used as spin filters[10,11]. EuO and EuS have been suggested as
spin filters by DiVincenzo[8], but the
compatibitity of these magnetic materials with conventional
semiconductors like GaAs or Si or Ge is unclear. The spin polarization in
tunneling current has exceeded $99 \%$ for EuO and EuS, and up to $90 \%$
in BeMnZnSe [11].

Here, we propose to use a narrow band QD array (QDA) weakly
coupled to leads as a spin filter. We find that electrons injected
from the leads into the QDA within a small window of the applied gate
voltage will favor a ferromagnetic state, as a result
of the strong electron correlation. In other words, the QDA
functions as a spin filter, which may be used to detect and
manipulate spin states. We also expect that the QDA has potential
applications in spintronics.

The device is described by the Hamiltonian,\\
 \( H =
\sum_{{\bf k},\sigma} \epsilon_{{\bf k}} a^{\dagger}_{{\bf k},
\sigma} a_{{\bf k},\sigma} +\sum_{{\bf p},\sigma} \epsilon_{{\bf
p}} b^{\dagger}_{{\bf p},\sigma}b_{{\bf p},\sigma}  +\sum_{i,{\bf
k},\sigma} V_{i,{\bf k}} a^{\dagger}_{{\bf k},\sigma} d_{i,\sigma}
+\sum_{i,{\bf p},\sigma} V_{i,{\bf p}} b^{\dagger}_{{\bf
p},\sigma} d_{i,\sigma}+ h.c. +\sum_{i,\sigma} E_0 n_{i,\sigma}
 + \sum_{i,j} t_{i,j} d^{\dagger}_{i,\sigma}
d_{j,\sigma} +\sum_{i,\sigma} U_i n_{i,\sigma}n_{i,-\sigma}, \)\\
where the first two terms describe the left lead and
right lead, respectively. The third and
fourth terms describe the coupling between the quantum dot (QD) and the two
leads. The fifth and sixth terms describe the energy level of the
quantum dot and interdot coupling. The last term describes the
intradot Coulomb interaction. We take into account only one
energy level for each dot and the nearest-neighbor coupling
between dots ($t_{i,j} = -t$ for nearest -neightbor $i,j$).  For
small size QDs, the energy difference between the ground state and
the first excited state is much larger than $t$ and the Coulomb
interaction $U$. Therefore, it is a good approximation to consider
just one energy level in each dot.

Using Keldysh's Green function method[12]
, we obtain the spin-dependent tunneling current
\begin{equation}
J_{\sigma} = -\frac{e}{\hbar}
\sum_{\bf k} [f_L-f_R] |V_{\bf k}|^2 ImG_{{\bf
k_{||} },\sigma}(\omega_{\bf k}),
\end{equation}
where $\omega_{\bf k}$ denotes the energy of the electron in the
leads with wave vector ${\bf k}$. $f_L=f(\omega_{\bf k} -\mu_L) $ and
$f_R =f(\omega_{\bf k} - \mu_R) $ are the Fermi
distribution function of the left lead and right lead,
respectively. $\mu_L$ and $\mu_R$ are the chemical potentials in the left
and right leads, respectively. They are related to the applied bias, $V_a$
by $\mu_L-\mu_R =eV_a$. For simplicity, we assume that the QD couples with
the left and right leads symmetrically, although it is
straight-forward to extend to the case with asymmetric coupling.
$V_{\bf k} = \sum_{j} V_{j,{\bf k}}e^{-i {\bf k_{||}}
\cdot R_j}$ where ${\bf k_{||}}$ is the projection of electron wave vector
${\bf k}$ in the QDA plane and $R_j$ is the position of the $j$-th QD.
We propose a setup in which a small bias $V_a$ ($eV_a$ is comparable to $t$)
is applied, and scan the gate voltage $V_g$ (which serves the purpose of
tuning the QD energy level $E_0$ relative to the Fermi level in the leads)
in order to observe the spin-dependent current.


The calculation of tunneling current is entirely determined by the
retarded Green function for the QDA. Finding the spin-polarized
Green function for the Hubbard model was considered a challenging
problem [13]. It is well known that the retarded Green function
obtained within the Hubbard approximation does not support any
magnetic ordering[14]. However, Harris and Lange[15] showed that
ferromagnetic ordering can be a stable state in the 3D Hubbard
model by introducing the spin dependent band shift. This mechanism
plays the crucial role for determining the ferromagnetic
state[15-18]. Beenen and Edwards[13] used the approach developed
by Roth[16] to study the 2D Hubbard model for the normal and
superconducting state of CuO$_2$. The quasi-particle excitation
energy they obtained is in very good agreement with quantum Monte
Carlo calculations[19]. Therefore we adopt Roth's procedure to
calculate the retarded Green function $G_{{\bf
k_{||}},\sigma}(\omega)$, while taking into account the coupling
between QDA and leads. In the weak-coupling limit ($t\ll U $), we
obtain
\begin{small}
\[ G_{{\bf k_{||}},\sigma}(\omega) =  \frac{1 -
n_{-\sigma}}{\omega - E_0- \epsilon_{k_{||}} (1 - n_{-\sigma}) -
n_{-\sigma} W_{{\bf k_{||}},-\sigma}+ i \Gamma({\bf k}) }\]
\[
 + \frac{n_{-\sigma}}{\omega - E_0 - U- \epsilon_{k_{||}}
n_{-\sigma} - (1- n_{-\sigma}) W_{{\bf k_{||}},-\sigma}+ i
\Gamma({\bf k}) }, \]\end{small}
where $\Gamma({\bf
k},\omega) $ denotes the the tunneling rate from the QDA to the leads
. It is cumbersome to fully include
the tunneling rates as a function of the wave vector and bias. We
treat $\Gamma$ as a constant parameter, even though it can be
determined via a numerical method [20]. This approximation is
valid for the small range of applied bais, because the potential
barrier between leads and QDA is high.
In the Coulomb blockade regime, it is adequate to consider the coupling
between the QDA and the leads within the Hatree-Fock
approximation[21](leading order). Here we consider a square
lattice with lattice constant $a$. The energy dispersion of
electrons in the QDA is then given by $\epsilon ({\bf k_{||}}) =
-2t [cos(k_x a)+ cos(k_y a)]$.
${\bf k_{||}}$ is restricted in the first Brillouin zone of the 2D lattice.
$ W_{{\bf k_{||}},-\sigma}$ denotes
the spin-dependent band shift, which is given by $n_{\sigma}
(1-n_{\sigma}) W_{{\bf k_{||}},\sigma} = w_{0,\sigma} +
w_{1,\sigma} \epsilon({\bf k_{||}})$, where $ w_{0,\sigma}$ denotes
the electron hopping correlation, while $w_{1,\sigma}$ consists
of three terms, which represent the density correlation,
spin correlation, and spin-flip correlation,
respectively[16]. Herrmann and Nolting [22] have proved that the
effect due to $w_{1,\sigma}$ is small for the ferromagnetic state
of a body central cubic lattice. This implies that the electron hopping
correction can maintain the stability of ferromagnetic state in 3D.
Nevertheless, $w_{1,\sigma}$ is kept in the present calculation in order
to obtain more accurate result.

When the chemical potential in the left lead is lower than $E_0 + U$ and
$ U\gg t$, the effect due to the high energy pole of the Green function
can be ignored.
Consequently, the infinite
U limit can be used in the calculation of $W_{{\bf k_{||}},-\sigma}$ [16]
which yields
\begin{small}
\[ n_{\sigma} W_{{\bf k_{||}},\sigma}
= \frac{-4t n_{1,\sigma}}{1-n_{\sigma}} - \epsilon ({\bf k_{||}})
\frac{n^2_{1,\sigma}(1-n_{\sigma}) + n_{1,\sigma}
n_{1,-\sigma}}{(1-n_{\sigma})(1-n_{\sigma} - n_{-\sigma})}
\]
with
\[ n_{1,\sigma} = \frac{-1}{2\pi} \sum_{{\bf k_{||}}} \frac{\epsilon
({\bf k})} {4t} \int d\omega [ f_L(\omega) + f_R(\omega)]
ImG_{{\bf k_{||}}, \sigma}(\omega). \]
\end{small}

The number of electron per dot is calculated by
\begin{small}
 \begin{equation} n_{\sigma} =
\frac{-1}{2\pi} \sum_{{\bf k_{||}}}\int d\omega [ f_L(\omega) +
f_R(\omega)] ImG_{{\bf k_{||}}, \sigma}(\omega).
\end{equation}
\end{small}
At zero temperature, the integral over $\omega$ can be carried out
analytically. Thus, we can obtain $n_{\sigma}$ and $n_{-\sigma}$
by solving two coupled one-dimensional integral equations self-consistently.
The parameters used in our calculations are
$E_0 = eV_g+(\mu_L +\mu_R)/2 +4 t$ (without applied bias),
$U = 20 t$,
$ a = 200 \AA$, and the effective mass of electrons in leads
(assumed to be GaAs) $m^* = 0.067 m_e$. 

The spin-dependent electron occupancy  $n_{\sigma} (\sigma=\uparrow,
\downarrow)$ as a function of the total
electron occupancy $n=n_{\uparrow}+n_{\downarrow}$ at zero
temperature for three different applied voltages is shown in Fig. 1;
solid line denotes $eV_a = 0.1 t$,
dashed line denotes $eV_a= 0.5 t$, and dotted line denotes
$eV_a = t$. We see a bifurcation for
$n_{\uparrow}$ and $n_{\downarrow}$ at $n=0.364, 0.367, 0.373$
for $eV_a = 0.1 t$, $0.5t$, and $t$, respectively, where the system
becomes spin polarized. The system remains ferromagnetic for $n$ within a
small window and reverts to the paramagnetic state at $n \approx 0.4$, beyond
which the pseudo-equilibrium
condition[16] can no longer be satisfied with $n_{\uparrow}
\neq n_{\downarrow}$.
As the applied bias is increased, the spin polarization
and the domain that maintains the spin polarization are reduced.
Eventually, the applied bias will totally destroy the spin polarization
of the system.

Fig. 2 shows the spin-dependent tunneling current as a function of
gate voltage ($V_g$) for various strengths of applied bias. We
define the spin polarization of the current as ${\cal P}_s =
(J_{\uparrow } -J_{\downarrow})/(J_{\uparrow} + J_{\downarrow })$.
We see that the maximun ${\cal P}_s = 0.4816, \; 0.3732,$ and
$0.3197 $ for $eV_a = 0.1, 0.5 t$ and $t$. This is much better
than the value (less than $1 \%$) achieved by using ferromagnetic
metals in contact with  semiconductors, because the conductivity
of metal is much higher than that of semiconductor.[23,24]
Although using magnetic semiconductor to replace the ferromagnetic
metal, the best value of ${\cal P}_s$ achieved is near $90
\%$[14], it remains to be seen if similar idea can be applied to
III-V and group-VI semiconductors. The spin dependent tunneling
current shown in Fig. 2 implies that we can readily manipulate the
spin polarization of the tunneling current by the gate voltage
without introducing magnetic field or magnetic dopants. Although
the value of the spin polarization ($P_s$) obtained here is not
very high, we believe that $P_s$ can be enhanced by using coupled
multiple layers of QDAs, since in the 3D Hubbard model the
ferromagnetic phase is stable over a wider range of $n$ and
$n_{\uparrow}$ (for majority carrier) can approach 1.[16]
According to Eq. (1), the tunneling current $J_{\downarrow}$ (for
minority carrier) can be reduced to zero as a result of the factor
$(1- n_{\uparrow})$, and $P_s$ can approach 1.


{\bf ACKNOWLEDGMENTS}

This work was supported by DARPA DAAD19-01-1-0324.


{\bf Figure Captions}

Fig. 1: Spin-dependent electron occupancy versus total occupancy $n$
for various strengths of applied bias; solid line ($eV_a = 0.1 t$),
dashed line ($eV_a = 0.5 t$), and dotted line ($eV_a = t $).

Fig. 2: Spin dependent tunneling current as a function of gate volatge
($V_g/4t$) for various strengths of applied bias; solid line
($eV_a = 0.1 t$), dashed line ($eV_a = 0.5 t$), and dotted line
($eV_a = t $).

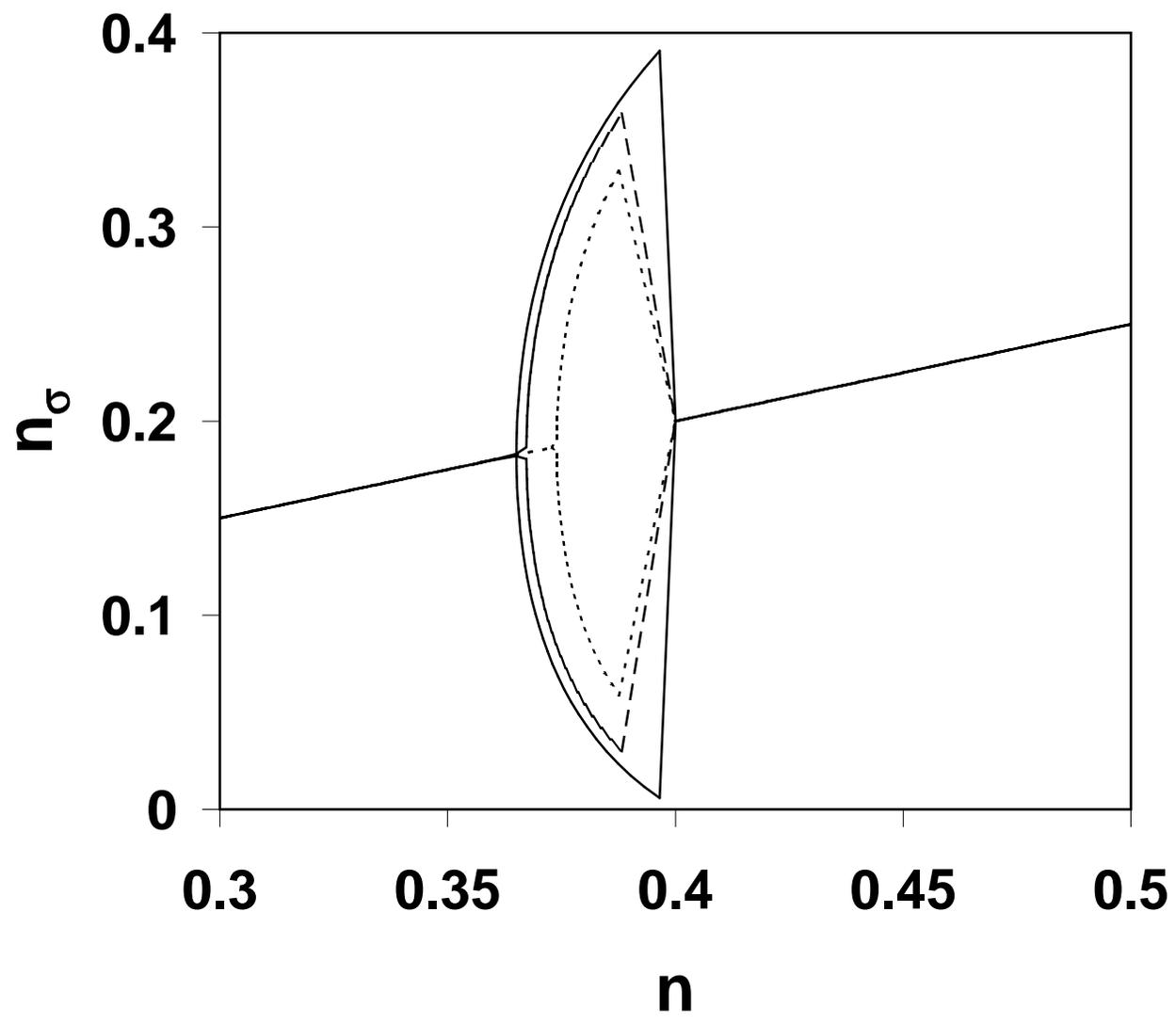
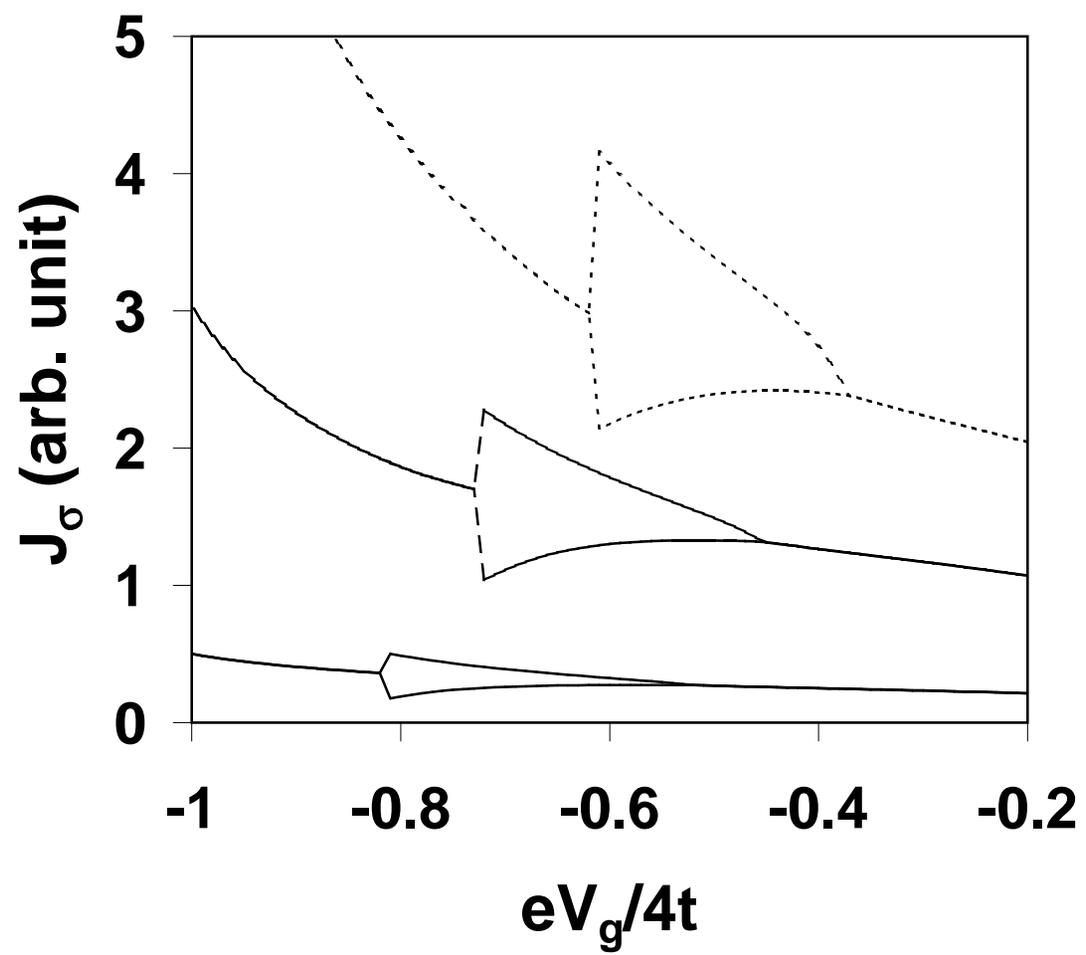
\end{document}